\documentclass{IEEEtran}
\usepackage{multirow}

\usepackage{graphicx}
\usepackage{subfigure}
\usepackage{url}
\usepackage{amsfonts}
\usepackage{amsmath}
\usepackage{bm} 
\usepackage{algorithm}
\usepackage{algorithmic}
\usepackage[table]{xcolor}
\let\labelindent\relax
\usepackage{enumitem}
\usepackage{comment}
\newcommand{\cmmnt}[1]{\ignorespaces}

\begin{document}
\title{Temporally Nonstationary Component Analysis; Application to Noninvasive Fetal Electrocardiogram Extraction}
\author{Fahimeh~Jamshidian-Tehrani, Reza~Sameni$^\textnormal{*}$,~\IEEEmembership{Senior~Member,~IEEE}, and Christian~Jutten,~\IEEEmembership{Fellow,~IEEE}%
\thanks{Manuscript received 22 February, 2019; revised 15 July, 2019, accepted 16 August 2019. \textit{Asterisk indicates corresponding author.}}%
\thanks{$^\textnormal{*}$F.~Jamshidian-Tehrani and R.~Sameni are with the School of Electrical \& Computer Engineering, Shiraz University, Shiraz, Iran. R.~Sameni and C.~Jutten are affiliated with GIPSA-lab, Universit\'e Grenoble Alpes, CNRS, Grenoble INP, 38000 Grenoble, France (e-mail: rsameni@shirazu.ac.ir).}%
}
\markboth{IEEE Transactions on Biomedical Engineering,  2020 [Authors' Preprint], Published Version's DOI: 10.1109/TBME.2019.2936943}{Jamshidian-Tehrani, Sameni and Jutten: Nonstationary Component Analysis}
\maketitle
\begin{abstract}
\textit{Objective:} Mixtures of temporally nonstationary signals are very common in biomedical applications. The nonstationarity of the source signals can be used as a discriminative property for signal separation. Herein, a semi-blind source separation algorithm is proposed for the extraction of temporally nonstationary components from linear multichannel mixtures of signals and noises. \textit{Methods:} A hypothesis test is proposed for the detection and fusion of temporally nonstationary events, by using ad hoc indexes for monitoring the first and second order statistics of the \textit{innovation process}. As proof of concept, the general framework is customized and tested over noninvasive fetal cardiac recordings acquired from the maternal abdomen, over publicly available datasets, using two types of nonstationarity detectors: 1) a local power variations detector, and 2) a model-deviations detector using the innovation process properties of an extended Kalman filter. \textit{Results:} The performance of the proposed method is assessed in presence of white and colored noise, in different signal-to-noise ratios. \textit{Conclusion and Significance:} The proposed scheme is general and it can be used for the extraction of nonstationary events and sample deviations from a presumed model in multivariate data, which is a recurrent problem in many machine learning applications.
\end{abstract}
\begin{keywords}
Nonstationary component analysis, semi-blind source separation, nonstationarity detection, generalized eigenvalue decomposition, approximate joint diagonalization, fetal electrocardiogram extraction.
\end{keywords}
\IEEEpeerreviewmaketitle
\section{Introduction}
\label{sec:introduction}

\PARstart{T}{he} problem of blind source separation (BSS) has been extensively studied in recent decades. Using minimal assumptions, such as the stochastic independence of the sources, powerful methods such as independent component analysis (ICA) have been developed for solving BSS \cite{ComonJutten2010handbook}. In real-world applications, additional priors commonly exist regarding the sources of interest. By incorporating these priors in source separation algorithms, the performance can be significantly improved, per application, as compared to generic source separation algorithms. Examples of such priors include temporal/spectral priors \cite{barros2001extraction, li2007sequential,yeredor2010second} and sparsity (in time, frequency or time-frequency domains) \cite{chenot2015robust}. In this context, \textit{temporal nonstationarity (TNS)} is one of the well-known priors used for source separation \cite{matsuoka1995neural, cardoso2001three,pham2001blind}. TNS can be in different stochastic attributes of the sources. In the second-order statistics case, a minimum number of two ``well-chosen'' covariance matrices, can be used in combination with generalized eigenvalue decomposition (GEVD) to accomplish source separation \cite{choi2002second}. Without accurate assumptions on time constants of TNS, the more common approach is the approximate joint diagonalization of more than two covariance (or lagged-covariance) matrices, which are calculated far enough in time, to capture the temporal nonstationarity of the sources \cite{pham2001blind}. A special class of nonstationary sources, with numerous biomedical applications corresponds to the case in which some sources have an on/off characteristic over time. In these cases, instead of a gradual change of the source statistics over time, nonstationary events are (irregularly) interleaved over time. The TNS may also be in the form of a \textit{rare event}. In these cases, typical global stochastic measures used for source separation--- calculated as averages over the entire data--- are commonly unaffected by such rare events and therefore the rare events remain inseparable. Biomedical examples of this phenomenon include: magnetic resonance artifacts in simultaneous electroencephalogram (EEG) recordings \cite{Amini08}; ocular EEG artifacts \cite{SameniGouyPailler2014}; event-related potentials in background EEG \cite{samadi2011identification}; biological signals corrupted by burst artifacts and device noise (e.g., due to loose electrode connections); fetal electrocardiogram (fECG) acquired from the maternal abdomen \cite{Sameni2008a}. In these examples, whenever a reliable reference channel or prior knowledge is available for detecting the TNS, classical BSS algorithms are successful in source separation. However, in many cases, the TNS epochs are not known a priori and are not visually detectable, due to the very low signal-to-noise ratio (SNR). For example, the detection of (single-trial) event related potentials, or noninvasive fECG extraction in low SNR. As a case study, we herein address this latter application, by combining nonstationary component analysis (NSCA) with \textit{local power envelope} variations and the \textit{innovation process} of an extended Kalman filter applied to the observations. It is known that the innovation process, in its broad sense, is the innovative part of the observations that cannot be estimated from prior knowledge and the data model. The deviation of the innovation process from its ideal properties (such as spectral whiteness) can be associated to factors that have changed the data model, which in our context correspond to TNS events. The proposed algorithm is shown to be applicable without assumptions such as the pseudo-periodicity of the maternal or fetal ECG, used in periodic component analysis ($\pi$CA) \cite{Sameni2008a}, which is useful for fECG extraction during irregular maternal or fetal beats. The performance of the proposed method is illustrated over a publicly available fECG database.

In Sections \ref{sec:background}, the theory of GEVD and NSCA are reviewed. The proposed TNS detection and fusion schemes are presented in Section \ref{sec:nonstationaritydetection}. Section \ref{sec:casestudy} presents the fECG extraction case study, followed by conclusions and future perspectives.

\section{Linear separation of temporally nonstationary sources}
\label{sec:background}

\subsection{Generalized eigenvalue decomposition}
\label{sec:GEVD}
For real symmetric matrices $\mathbf{A},\mathbf{B} \in \mathbb{R}^{n \times n}$, \textit{generalized eigenvalue decomposition (GEVD)} of {the matrix pair} $(\mathbf{A},\mathbf{B})$ consists of finding $\mathbf{W} \in \mathbb{R}^{n \times n}$ and $\bm{\Lambda}\in \mathbb{R}^{n \times n}$, such that
\begin{equation}\label{eq:EVD}
		\mathbf{W}^T \mathbf{A} \mathbf{W} = \bm{\Lambda}, \quad \mathbf{W}^T \mathbf{B} \mathbf{W} = \mathbf{I}_n
\end{equation}
where $\bm{\Lambda} = \textbf{\text{diag}}(\lambda_1, \ldots, \lambda_n)$ contains the generalized eigenvalues, corresponding to the eigenmatrix $\mathbf{W}=[\textbf{w}_1,\ldots,\textbf{w}_n]$, with real eigenvalues sorted in ascending order on its diagonal. Symmetric positive definite matrix pairs have real positive eigenvalues and the first eigenvector $\textbf{w}_1$ maximizes the \textit{Rayleigh quotient} \cite{Strang1988}:
\begin{equation}\label{eq:Rayleigh}
J(\textbf{w}) = \frac{\textbf{w}^T \mathbf{A} \textbf{w}}{\textbf{w}^T \mathbf{B} \textbf{w}}
\end{equation}
It can be shown that all ICA methods based on pre-whitening can be eventually converted into a GEVD problem of two (problem-specific) matrices \cite{Sameni2008a}. Therefore, in semi-blind source separation problems, in which prior knowledge regarding the underlying components exists, the problem of source separation can be considered as a \textit{matrix design problem}. The performance of GEVD-based source separation and generic methods for choosing the proper matrix pair have been addressed in previous research \cite{yeredor2011performance,yeredor2009optimal}. Herein, we adopt a prior-based approach for the selection of the target matrices.
\subsection{Nonstationary component analysis (NSCA)}
\label{sec:NSCA}
Let us consider multivariate data $\mathbf{x}(t) \in \mathbb{R}^n$ ($t\in \mathcal{T}$), where $\mathcal{T}$ denotes the entire set of available discrete-time samples and $\mathcal{P} \subset \mathcal{T}$ is a subset of these samples, which are considered as being \textit{nonstationary} or \textit{odd events} that do not follow the (average) background model in certain aspects. For example, they can correspond to outliers with statistical properties that differ from the other samples. By defining \textit{ad hoc} (problem-specific) measures of signal nonstationarity, one can perform a sample-wise \textit{hypothesis test} for the identification of the TNS:
\begin{equation}
\begin{array}{rl}
         \mathcal{H}_0: t \notin \mathcal{P} \\
         \mathcal{H}_1: t \in \mathcal{P}
    \end{array}
\label{eq:hypothesistestGeneral}
\end{equation}
Denoting the subset of samples that satisfy the alternative hypothesis $\mathcal{H}_1$ with $u \in \mathcal{P}$, a special case of GEVD is obtained by finding the matrix $\mathbf{W}$, which satisfies (\ref{eq:EVD}), for $\mathbf{A} = \mathbb{E}_u\{\mathbf{x}(u)\mathbf{x}(u)^T\}$, $\mathbf{B} = \mathbb{E}_t\{\mathbf{x}(t) \mathbf{x}(t)^T\}$, where $\mathbb{E}_t\{\cdot\}$ and $\mathbb{E}_u\{\cdot\}$ denote averaging over all time samples $t\in \mathcal{T}$ and $u\in \mathcal{P}$, respectively. Using this matrix, the linear transform $\mathbf{y}(t) = \mathbf{W}^T \mathbf{x}(t)$ extracts $n$ uncorrelated channels with maximal energy over the subset of time samples $\mathcal{P}$. In other words, $\mathbf{W}$ retrieves uncorrelated linear mixtures of $\mathbf{x}(t)$ with maximal energy during the TNS epochs.

\subsection{Joint approximate diagonalization of multiple matrices}
While the exact diagonalization of two symmetric matrices is always possible by GEVD, no more than two matrices can be exactly diagonalized using a single {linear} transform, unless if the matrices belong to the same eigenspace \cite{Strang1988}--- a condition which does not necessarily hold in the problem of interest. Therefore, methods based on \textit{approximate joint diagonalization} (AJD) of more than two matrices {have been developed} \cite{Cardoso1993,pham2001joint,tichavsky2009fast}.

In the problem of interest, if there is evidence of TNS over different subsets of time samples, one can apply AJD to {a} set of {$N$} matrices
\begin{equation}
\mathbf{C}_i = \mathbb{E}_{u_i}\{\mathbf{x}(u_i) \mathbf{x}(u_i)^T\}, \quad i = 1, \ldots, N
\label{eq:multiplecovs}
\end{equation}
where {$u_i\in \mathcal{P}$ are} subsets of time samples corresponding to different nonstationary events. This approach is commonly more robust to covariance matrix estimation errors, as compared with GEVD.

\section{Automatic nonstationarity detection}
\label{sec:nonstationaritydetection}
Nonstationarity, generally refers to variations in signal attributes in time, frequency, time-frequency, stochastic properties, etc. Herein, we focus on two such attributes: (1) nonstationary stochastic processes with time-variant power envelopes; (2) nonstationary stochastic processes in presence of a dominant background signal, which has a known stochastic behavior represented by a dynamic model. The criteria used for identifying TNS epochs based on these attributes are presented in the sequel. The significance of these attributes are shown in the case study of Section \ref{sec:casestudy}.

\subsection{The local power envelope (LPE)}
\label{sec:localenergy}
Consider the time-varying power of a signal $s(t)$ over a sliding window of length $w$:
\begin{equation}
P_w(t) = \displaystyle \frac{1}{w}\sum_{a = -\frac{w}{2}}^{\frac{w}{2}} |s(t-a)|^2
\label{eq:TVEnergy}
\end{equation}
The ratio of $P_w(t)$ for two windows of lengths $w= w_1$ and $w=w_2$ ($w_2 \gg w_1$) can be used as a measure for detecting fast local nonstationary epochs within a slowly varying (or stationary) background activity:
\begin{equation}
\rho(t) = \displaystyle \frac{P_{w_1}(t)}{P_{w_2}(t)}
\label{eq:TVEnergyRatio}
\end{equation}
which we name the \textit{local power envelope (LPE)}. For a global measure, the denominator {$P_{w_2}(t)$} can be replaced with the average signal power $P_\infty$. The values of $\rho(t)$ significantly smaller or larger than one correspond to time epochs that are different (nonstationary) from the background activity. The rationale behind the above definition is that a stationary signal has a consistent energy profile over time and notable deviations of the LPE from unity (with application-dependent window lengths $w_1$ and $w_2$) are indicators of nonstationary epochs.

The LPE can be used to extract the TNS epochs of a signal as follows:
\begin{equation}
\theta_{k}^{(\text{LPE})} = \{t\quad\!\!\!|\quad\!\!\!  {\rho_k(t) \geq \zeta_k^u \text{ or } \rho_k(t) \leq \zeta_k^l}, t\in \mathcal{T}\}
\label{eq:localenergyindex}
\end{equation}
where $\zeta_k^u$ and $\zeta_k^l$ are predefined upper and lower thresholds satisfying $\zeta_k^u > 1 > \zeta_k^l \geq 0$ and $\rho_k(t)$ is the LPE calculated for the $k$th channel of a given signal.

\subsection{The innovation process of an extended Kalman filter}
\label{sec:KalmanFilterReview}
In many applications, the TNS of interest is dominated by other background activities, avoiding the visual/automatic detection of the nonstationary process and even without having a notable footprint in the signal's LPE. For applications in which the background activity has known stochastic properties, which could be modeled with a linear or nonlinear dynamic model, the innovation process of a Kalman or extended Kalman filter (EKF) can be used to detect the nonstationary events.

Consider the general state-space representation of a discrete-time stochastic process:
\begin{equation}
\label{eq:NLDynamic}
  \begin{array}{rll}
  \mathbf{s}(t+1) & \!\!\!= \displaystyle\mathbf{f}\left(\mathbf{s}(t), \bm{w}(t), t\right) & \!\!\!\approx \mathbf{A}(t) \mathbf{s}(t) + \tilde{\bm{w}}(t)\\
  \mathbf{x}(t) & \!\!\!= \displaystyle\mathbf{g}\left(\mathbf{s}(t), \bm{\eta}(t), t\right) + \mathbf{n}(t) & \!\!\!\approx \mathbf{H}(t) \mathbf{s}(t) + \tilde{\bm{v}}(t)
  \end{array}
\end{equation}
where $\mathbf{x}(t)$ are multichannel \textit{observations} (\textit{measurements}), $\mathbf{s}(t)$ denotes the \textit{state vector}, and $\mathbf{n}(t)$ is the \textit{nonstationary component} that we seek to detect and only exists during  {the subset of} time instants {$\mathcal{P} \subset \mathcal{T}$}. The right-side equations of (\ref{eq:NLDynamic}) are linear approximations of the generally nonlinear {state and observation functions} $\mathbf{f}(\cdot)$ and $\mathbf{g}(\cdot)$. The vector $\bm{w}(t)$ denotes zero-mean white \textit{process noise} and $\tilde{\bm{w}}(t)$ is the process noise after linearization. The vector $\bm{\eta}(t)$ denotes the \textit{measurement noise} and $\tilde{\bm{v}}(t) = \tilde{\bm{\eta}}(t) + \mathbf{n}(t)$ is the superposition of the measurement noise vector after linearization $\tilde{\bm{\eta}}(t)$ and the nonstationary component $\mathbf{n}(t)$, which may be present or absent during different time epochs. We further define $\mathbf{Q}(t)=\mathbb{E}\{\tilde{\bm{w}}(t)\tilde{\bm{w}}(t)^T\}$ and $\mathbf{R}(t)=\mathbb{E}\{\tilde{\bm{\eta}}(t)\tilde{\bm{\eta}}(t)^T\}$, the covariance matrices of the (zero-mean) process and observation noises after linearization, respectively. The ultimate objective is to detect the time instants in which the nonstationary component $\mathbf{n}(t)$ exists, using a {sample-wise}
hypothesis test:
\begin{equation}
    \begin{array}{rl}
         \mathcal{H}_0: & \tilde{\bm{v}}(t) = \hat{\bm{\eta}}(t)\\
         \mathcal{H}_1: & \tilde{\bm{v}}(t) = \hat{\bm{\eta}}(t) + \mathbf{n}(t)
    \end{array}
\label{eq:hypothesistest}
\end{equation}
The above test can be more generally regarded as a means of detecting whether or not a signal sample (or signal epoch) fulfills the presumed data model.

{Herein, we propose to use the EKF \textit{innovation process}} to solve this hypothesis test. The innovation process is defined:
\begin{equation}
	\label{eq:innovation}
	\hat{\bm{v}}(t) \stackrel{\Delta}{=} \mathbf{x}(t) - \hat{\mathbf{x}}(t)
\end{equation}
where $\hat{\mathbf{x}}(t)$ is an estimate of the observation $\mathbf{x}(t)$ before its arrival, obtained from standard {Kalman filter} equations \cite{GrewalAndrews01}. The innovation process is in fact an estimate of the measurement (observation) noise \textit{before} the arrival of the $t$th measurement, using the system's dynamics and measurements up to sample $t-1$. If the parameters of a {linear} Kalman filter are correctly set, in absence of $\mathbf{n}(t)$, i.e., the null hypothesis $\mathcal{H}_0$, the innovation process has the following properties:
\begin{enumerate}[label= P{{\arabic*}})]
\item $\mathbb{E}\{\hat{\bm{v}}(t)\} = \mathbf{0}$,
\item $\bm{\Gamma}(t) \stackrel{\Delta}{=} \mathbb{E}\{\hat{\bm{v}}(t) \hat{\bm{v}}(t) ^ T\} = \mathbf{H}(t) \hat{\mathbf{P}}(t) \mathbf{H}(t)^T + \mathbf{R}(t)$,
\item $\mathbb{E}\{\hat{\bm{v}}(t) \hat{\bm{v}}(t') ^ T\} = \mathbf{0}$ for $t \neq t'$
\end{enumerate}
where $\hat{\mathbf{P}}(t)$ is the covariance matrix of the state vector estimate before the arrival of the $t$th measurement. The first two properties assure that the Kalman filter's estimate of the measurement noise is in accordance with the presumed model parameters. The third property guarantees that the Kalman filter has successfully estimated all the predictable parts of the observations, leaving white noise as its residue. The violation of any of these properties is an indication of parameter mis-selection (e.g., in $\mathbf{Q}(t)$ or $\mathbf{R}(t)$), or a model mismatch due to the presence of $\mathbf{n}(t)$, i.e., the alternative hypothesis $\mathcal{H}_1$. Although the above properties originally belong to the linear Kalman filter, for detection purposes--- as in our case--- in which the primary interest is event detection (rather than optimal filtering), the above properties can be equally used to monitor the well-functioning of the EKF. For this, we introduce three indexes to monitor the expected innovation process properties:

\subsubsection{Innovation process mean}
\label{sec:innovationmean}
For a sliding window of length $w_a$, we define
\begin{equation}
a_k(t) = \frac{1}{w_a}\displaystyle \sum_{s = t - \frac{w_a}{2}}^{t + \frac{w_a}{2} - 1} \hat{v}_k(s), \quad k = 1,\ldots, n.
\label{eq:innovationmean}
\end{equation}
where $\hat{v}_k(t)$ is the $k$th {channel} of $\hat{\bm{v}}(t)$. Property (P1) implies that $a_k(t)$ should be close to zero. By comparing its absolute value with a predefined threshold $\mu_k$, the mean nonstationary epochs of the measurements are estimated\footnote{For a causal implementation, (\ref{eq:innovationmean}) can be shifted by $w_a/2$ samples.}:
\begin{equation}
\theta_{k}^{(m)} = \{t\quad\!\!\!|\quad\!\!\!|a_k(t)| \geq \mu_k, t\in \mathcal{T} \}
\label{eq:innovationmeanindex}
\end{equation}
which denotes the TNS instants for the $k$th channel (the superscript $m$ denotes mean-nonstationary). For multichannel observations, $\mathbf{a}(t)=[a_1(t), \ldots, a_n(t)]$ is a vector compared with the threshold vector $\bm\mu = [\mu_1, \ldots, \mu_n]$. In the sequel, the time instants for which (\ref{eq:innovationmeanindex}) holds are called the \textit{mean-nonstationary} epochs of the signal.

\subsubsection{Innovation process variance}
\label{sec:innovationvariance}
According to (P2), for a well-functioning Kalman filter, during the null hypothesis $\mathcal{H}_0$, the following index, which corresponds to the average ratio {between} the actual {and} presumed observation noise variances, should be close to one:
\begin{equation}
\gamma_k(t) = \frac{1}{w}\displaystyle \sum_{s = t - \frac{w}{2}}^{t + \frac{w}{2} - 1} \frac{[\hat{v}_k(s) - a_k(s)]^2}{\Gamma_{kk}(s)}
\label{eq:innovationvariance}
\end{equation}
where $\hat{v}_k(t)$ and $a_k(t)$ are respectively the $k$th channel of $\hat{\bm{v}}(t)$ and $\mathbf{a}(t)$ (calculated from the measurements), and $\Gamma_{kk}(t)$ is the $k$th diagonal entry of the innovation process covariance matrix $\bm\Gamma(t)$ (calculated from the presumed EKF covariance matrices). Although the off-diagonal entries of the innovation process covariance matrix are also informative for nonstationarity detection, for proof of concept, we only consider the diagonal entries, which correspond to the channel-wise innovation process variances. Next, by comparing $\gamma_k(t)$ with predefined (channel-dependent) upper and lower thresholds $\lambda_k^u$ and $\lambda_k^l$, which satisfy $\lambda_k^u \geq 1 > \lambda_k^l \geq 0$, the TNS epochs of the measurements corresponding to dynamic changes in the signal or model parameters are detected:
\begin{equation}
\theta_{k}^{(v)} = \{t\quad\!\!\!|\quad\!\!\! {\gamma_k(t) \geq \lambda_k^u \text{ or }  \gamma_k(t) \leq \lambda_k^l} , t\in \mathcal{T} \} 
\label{eq:innovationvarianceindex}
\end{equation}
where the superscript $v$ denotes variance nonstationarity and we refer to the time instants for which (\ref{eq:innovationvarianceindex}) holds as the \textit{variance-nonstationary} epochs of the signal.

\subsubsection{Innovation process whiteness}
\label{sec:innovationcolor}
In order to assess (P3), we define an estimate of the cross covariance matrix of the innovation signal over a sliding window of length $w_r$:
\begin{equation}
\mathbf{r}(t, \tau) = 
\displaystyle \frac{1}{w_r}\sum_{s = t - \frac{w_r}{2}}^{t+\frac{w_r}{2}-1} {\bm{\vartheta}}(s -\frac{\tau}{2}){\bm{\vartheta}}(s + \frac{\tau}{2})^T
\label{eq:innovationwhitenessindex}
\end{equation}
where ${\bm{\vartheta}}(t)\stackrel{\Delta}{=}\hat{\bm{v}}(t) - \mathbf{a}(t)$. According to (P3), we should {ideally} have $\mathbf{r}(t, \tau) = \bm{r}(t)\bm{\delta}(\tau)$, corresponding to white noise. However, in practice due to the nonstationary events, the spectrum may become colored, expanding $\mathbf{r}(t, \tau)$ along $\tau$. In order to quantify the spectral color, for each $t$ and each channel $k$, a two-parameter function of the form $q_k(t)\exp({\frac{-|\tau|}{\epsilon_k(t)}})$ can be empirically fitted over each diagonal entry of $\mathbf{r}(t, \tau)$, by \textit{nonlinear least square error} fitting. The two model parameters $q_k(t)$ and $\epsilon_k(t)$ are next tracked over time, to identify nonstationary epochs of the innovation process. Mathematically,
\begin{equation}
    [q_k^*(t), \epsilon_k^*(t)] =  \underset{q_k(t), \epsilon_k(t)}{\operatorname{\arg\min}} \mathbb{E}_\tau\{|r_{kk}(t, \tau) - q_k(t)\exp({\frac{-|\tau|}{\epsilon_k(t)}})|^2\}
\end{equation}
Finally, by monitoring $q_k(t)$ and $\epsilon_k(t)$ with respect to predefined thresholds $\xi_k$ and $\kappa_k$, the nonstationary epochs of the measurements are detected:
\begin{equation}
\begin{array}{l}
	\theta_{k}^{(w_q)} = \{t\quad\!\!\!|\quad\!\!\!|q_k(t)| \geq \xi_k, t\in \mathcal{T} \}\\
	\theta_{k}^{(w_{\epsilon})} = \{t\quad\!\!\!|\quad\!\!\!|\epsilon_k(t)| \geq \kappa_k, t\in \mathcal{T} \}
\end{array}
\label{eq:innovationwhitenessindexes}  
\end{equation}
We name the time instants for which (\ref{eq:innovationwhitenessindexes}) holds, the \textit{spectrum-nonstationary} points of the signal. Note that various schemes can be used for spectral color tracking. For instance, one may fit a time-varying autoregressive (TVAR) model of fixed-order over $\hat{\bm{v}}(t)$ and track the DC gain and pole location(s) of this model over time, or may additionally monitor the non-diagonal entries of $\mathbf{r}(t, \tau)$ for inter-channel cross-correlations.

\subsection{Nonstationarity label fusion}
\label{sec:fusion}
In Section \ref{sec:KalmanFilterReview}, various indexes were proposed for detecting nonstationary epochs of multichannel data. Before utilizing the detected epochs in the NSCA algorithm, one may incorporate additional priors (if available) to correct the detected time labels. Examples of such corrections include:
\begin{itemize}[leftmargin=*]
\item \textit{Pruning/Insertion}: {Certain time instants may be added to or eliminated from the estimated point set, using additional priors.} For example, in the maternal-fetal ECG application studied in Section \ref{sec:casestudy}, the maternal and fetal QRS peaks may frequently overlap in time (due to their asynchronous heart-rates). Therefore, one may exclude the time instants of the maternal QRS that overlap with the fetal QRS nonstationary windows, to permit the separation of the maternal and fetal ECG.
\item \textit{Union}: The TNS points obtained from several indexes can be merged. This is a useful operation for combining different measures of nonstationarity or nonstationarities detected from multiple channels.
\item \textit{Intersection}: Taking the intersection of the TNS points between several indexes is useful for detecting TNS epochs that are common between multiple detection indexes or channels.
\item \textit{Voting}: By voting between the {TNS} point candidates extracted from different channels/indexes, we obtain TNS points by consensus between multiple indexes or channels. The voting mechanism can be weighted in favor of the most reliable channels/indexes.
\end{itemize}
Examples of applying these operators for {TNS} time label corrections is shown in the case study of Section \ref{sec:casestudy}.

\section{Case study: noninvasive fetal ECG extraction}
\label{sec:casestudy}
\subsection{Motivation}
The problem of fetal electrocardiogram (fECG) extraction from multichannel noninvasive maternal abdominal recordings is a classical application for blind and semi-blind source separation algorithms \cite{Car98,LMV00,Vrins04a,Sameni2008a,Sameni2008}. While generic and problem specific BSS and semi-BSS methods have been developed and applied for fECG extraction, the problem remains a challenge in low SNR and for abnormal maternal/fetal ECG \cite{SameniClifford2010}. Some of the major challenges and open problems in this domain were recently addressed in \cite{jamshidian2018fetal}. The most recent advances in this field are semi-BSS algorithms that presume pseudo-periodicity of the maternal/fetal ECG \cite{jamshidian2018fetal}, which is a limiting assumption in practice. In fact, such semi-BSS methods commonly require the R-peak locations of the mother and/or fetus for a robust performance. However, if the maternal/fetal ECG are highly irregular, the fetus moves, or the R-peak estimates are inaccurate (e.g., due to low SNR or missing R-peaks), the performance of these methods degrades.

As compared to pseudo-periodicity, a relaxing assumption, which holds for both regular and irregular beats, is that the fetal {QRS has a bumpy shape, which can be distinguished from the background noise (not necessarily by visual inspection, and sometimes requiring signal processing)}, and that the maternal ECG (mECG) can be quite accurately modeled with a dynamic model{, with its R-peaks rather easily detectable from an independent chest lead (or even from certain abdominal leads, depending on the lead configuration)}. This idea is used in the sequel to develop an NSCA customized for fECG extractions.

\subsection{FECG detection from Kalman filter innovation process}
Various dynamic models have been proposed for ECG modeling. For proof of concept, we use the modified polar version of the McSharry-Clifford ECG model \cite{McSharry2003,SSJC06}, which can be integrated in the proposed scheme.

Let us assume that the maternal abdominal signals consist of the mECG $s_m(t)$, the fECG $s_f(t)$ and background noise $\nu(t)$. Using the nonlinear state-space model proposed in \cite{SSJC06} for mECG modeling, the following set of equations can be written for the maternal body surface recorded signals $x(t)$:

\noindent\textit{Process equations}:
\begin{equation}
  \begin{array}{l}
  \psi(t+1)=[\psi(t) + \omega_m(t)]\mod(2\pi)\\
  \resizebox{.96\hsize}{!}{$s_m(t+1)=s_m(t)\displaystyle-\omega_m(t)\sum_{i=1}^K\frac{\alpha_i \tilde{\psi}_i(t)}{b_i^2} \exp(\frac{-\tilde{\psi}_i(t)^2}{2b_i^2})+w(t)$}
  \end{array}
\label{eq:ECGKFProcess}  
\end{equation}

\noindent\textit{Observation equations}:
\begin{equation}
	\begin{array}{l}
		\phi(t) = \psi(t) + \nu(t)\\
		x(t) = s_m(t) + s_f(t) + \eta(t)
	\end{array}
\label{eq:ECGKFObservation}  
\end{equation}
where $\tilde{\psi}_i(t)=[\psi(t)-\psi_i]\mod(2\pi)$, $\omega_m(t) = 2\pi f_m(t)/f_s$ is the beat-wise maternal heart-rate normalized angular velocity, $f_m(t)$ is the beat-wise maternal heart-rate in Hertz, $f_s$ is the sampling frequency in Hertz, $\alpha_i$, $b_i$, and $\psi_i$ are the amplitude, width and center parameters of the $i$th Gaussian kernel, and $K$ is the number of Gaussian kernels used for modeling the {mECG} morphology. In (\ref{eq:ECGKFProcess}) and (\ref{eq:ECGKFObservation}), $\psi(t)$ and $s_m(t)$ are the state variables; $\phi(t)$ is the cardiac phase measurement (obtained by maternal R-R interval calculation and a linear phase map); $x(t)$ is the maternal abdominal ECG measurement; $w(t)$ denotes the process noise; $\nu(t)$ is the phase measurement noise and $\eta(t)$ is the ECG measurement noise. Further details regarding this model and its parameters can be followed from \cite{SSJC06,SSJ08}. As suggested in \cite{SSJC06}, this model can be used in an EKF for estimating the mECG $\hat{s}_m(t)$. At the same time, according to the details in Section \ref{sec:KalmanFilterReview}, the innovation process $x(t) - \hat{s}_m(t)$ is an estimate of $s_f(t) + \eta(t)$. In absence of the fECG peaks, the innovation process should be white noise with the aforementioned properties (P1-P3). However, under the fetal QRS, the statistical properties of the innovation process significantly change, which can be detected using the indexes proposed in Sections \ref{sec:innovationmean}, \ref{sec:innovationvariance}, and \ref{sec:innovationcolor}. Based on this idea, for $n$-channel noninvasive maternal abdominal recordings $\mathbf{x}(t)$, a realization of the proposed NSCA algorithm is detailed in Algorithm \ref{alg:NSCAFetal}.

\begin{algorithm}
\caption{Noninvasive fetal ECG extraction {by} nonstationary component analysis \label{alg:NSCAFetal}}
\begin{algorithmic}[1]
\REQUIRE{Maternal abdominal recordings $\mathbf{x}(t) \in \mathbb{R}^n$}
\STATE Find the maternal R-peaks from an arbitrary maternal abdominal lead or a reference chest lead.
\STATE Calculate the average mECG beat from all channels by synchronous averaging or robust weighted averaging \cite{Leski2004}.
\STATE Fit a sum-of-Gaussians model over the average mECG beat and {estimate} the Gaussian model parameters by nonlinear least squares error estimation \cite{SSJC06}.
\STATE Apply an EKF to recover the mECG $\hat{\mathbf{m}}(t)$ and calculate $\hat{\bm{v}}(t)$, the innovation signal of all channels \cite{SSJC06}.
\STATE Calculate {$\mathcal{P}=\{\theta_{k}^{(m)}, \theta_{k}^{(v)}, \theta_{k}^{(w_{\epsilon})}, \theta_{k}^{(w_q)}, \theta_{k}^{(\text{LPE})} \}$, i.e., the subset of TNS epochs} over each channel {$k$}, using the indexes proposed in Section~\ref{sec:nonstationaritydetection}.
\STATE Calculate $\mathbf{C}_x$, the covariance matrix of $\mathbf{x}(t)$.
\STATE Calculate {$\{\mathbf{C}_1, \ldots, \mathbf{C}_N\}$}, the covariance matrices of the nonstationary epochs according to (\ref{eq:multiplecovs}).
\STATE Find $\mathbf{W}\in\mathbb{R}^{n\times n}$, which diagonalizes $\mathbf{C}_x$ and approximately diagonalizes $\{\mathbf{C}_1, \ldots, \mathbf{C}{_N}\}$ by AJD$^{*}$.\label{algostep:jade} 
\STATE Calculate $\mathbf{y}(t) = \mathbf{W}^T \mathbf{x}(t)$. The fECG are expected to be extracted among the channels of $\mathbf{y}(t)$ (among the first few channels for the GEVD version of the algorithm).
\end{algorithmic}
\rule{\columnwidth}{0.5pt}
\textbf{Note:} For a GEVD implementation, step (\ref{algostep:jade}) can be replaced by GEVD of the matrix pair $(\mathbf{C}_x,\mathbf{C}_\theta)${, where} the covariance matrix $\mathbf{C}_\theta$ can be calculated from an arbitrary channel {TNS} epoch, or the \textit{union} of the TNS epochs of all channels $\theta = \theta_1 \cup \ldots \cup \theta_n$ (for an inclusive nonstationary component search), or over the \textit{intersection} of the TNS epochs of different channels $\theta = \theta_1 \cap \ldots \cap \theta_n$ (resulting in less susceptibility to channel-wise outliers). Refer to Section~\ref{sec:fusion} for further details.
\end{algorithm}

\subsection{Results}
\label{sec:results}
\begin{figure}[tb]
\centering
\includegraphics[width=.48\textwidth]{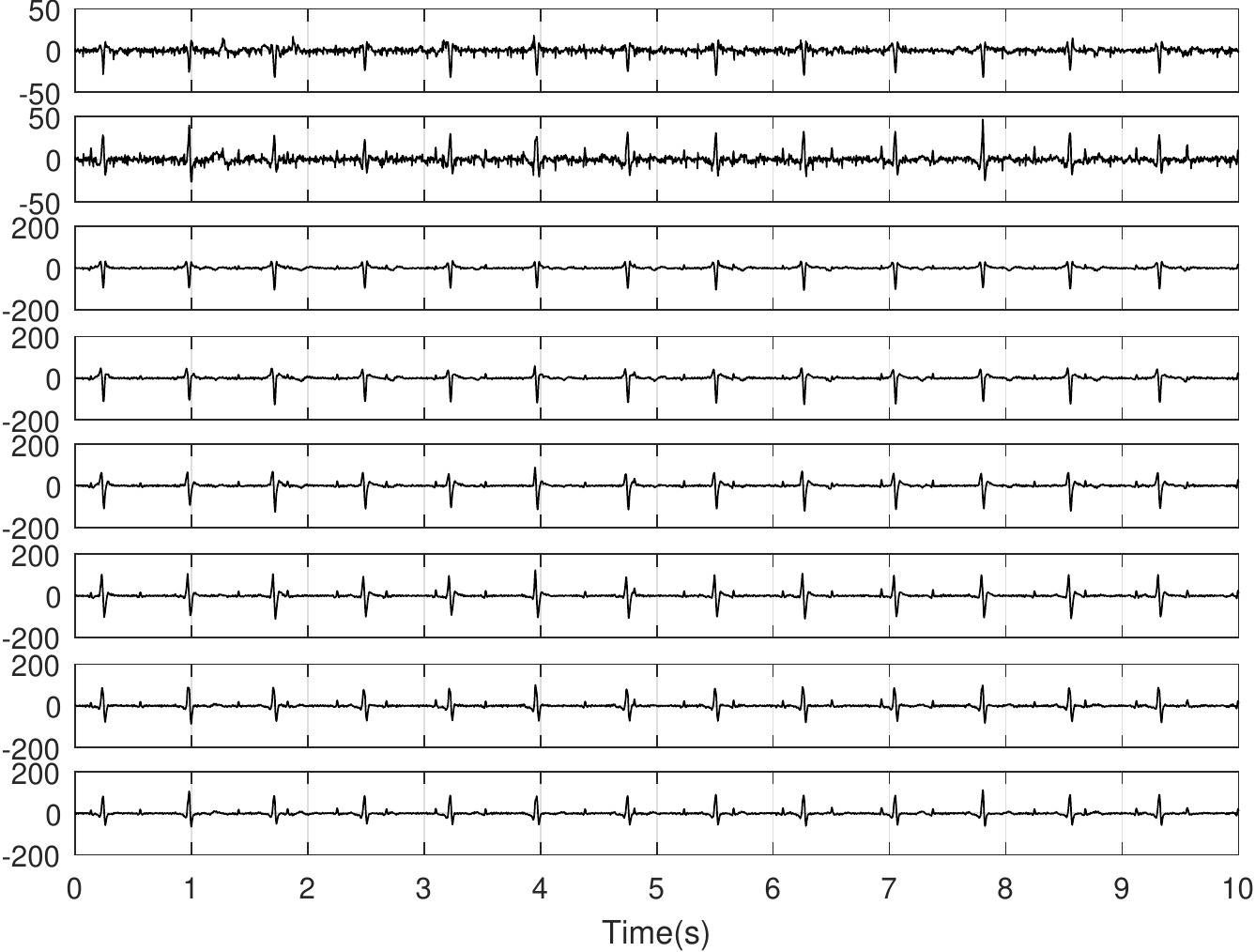}
\caption{A sample 8-channel maternal abdominal record.}
\label{fig:original}
\end{figure}

The proposed method is evaluated over the online available {fECG} dataset of the \textit{open-source electrophysiological toolbox (OSET)} \cite{OSET3.14}. For a visual illustration, we use the test sample \texttt{signal\_19} of this dataset, which has been recorded from eight maternal abdominal channels with a sampling rate of 500~Hz, using a system developed at the National Aerospace University, Kharkov, Ukraine. Ten seconds of this sample data is shown in Fig.~\ref{fig:original}.

\begin{figure}[tb]
\centering
\includegraphics[width=.48\textwidth]{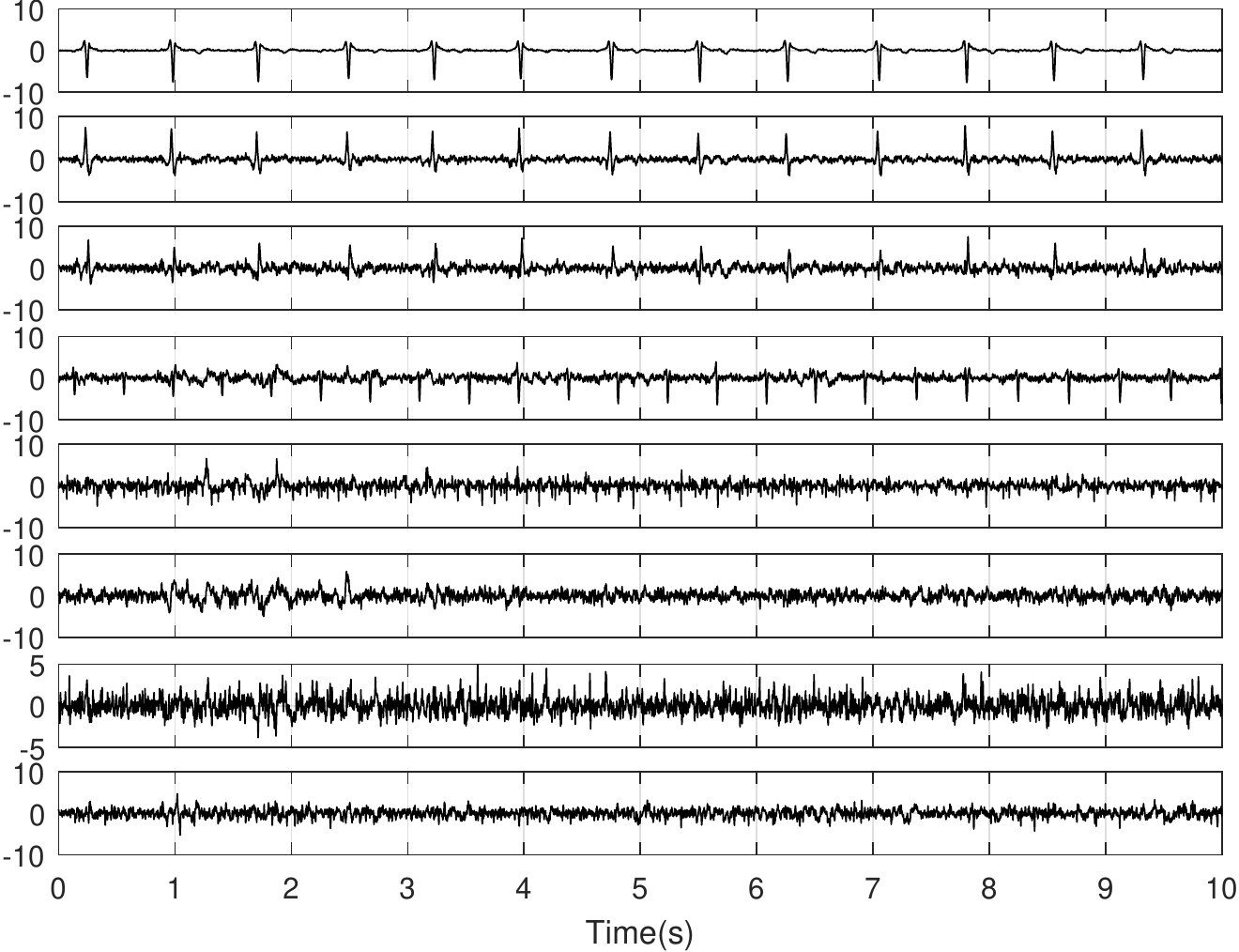}
\caption{Result of applying JADE on the dataset of Fig.~\ref{fig:original}. Components 1{, 2 and 3} correspond to the mECG and component {4 corresponds} to the fECG.}
\label{fig:ICA}
\end{figure}

\begin{figure}[tbh]
\centering
\includegraphics[width=.48\textwidth]{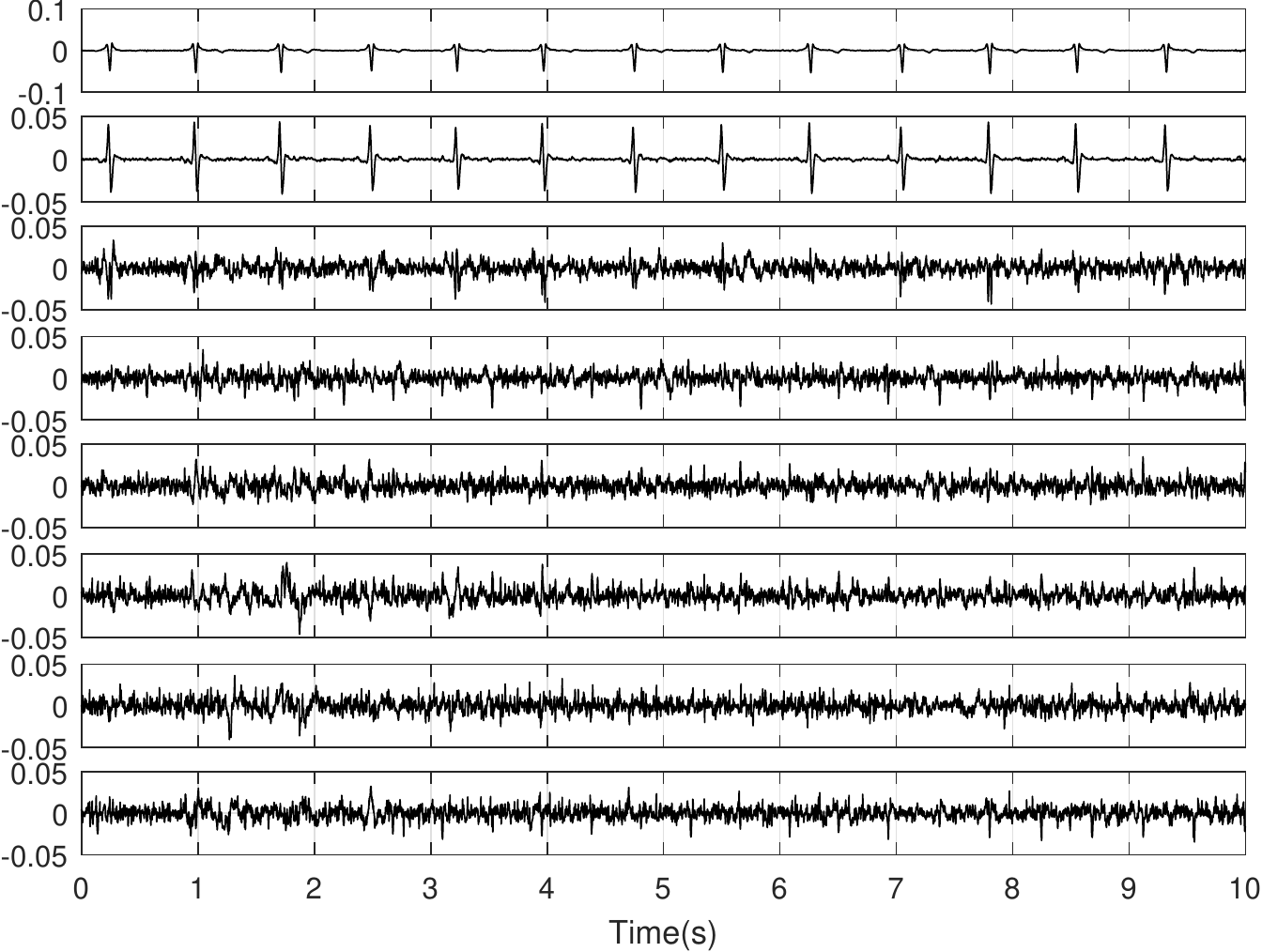}
\caption{Result of applying $\pi$CA on the dataset of Fig.~\ref{fig:original} using the maternal R-peaks. In this algorithm, the components are automatically ranked from top to bottom according to their similarity to the maternal ECG.}
\label{fig:PiCA}
\end{figure}

For comparison, the results of applying the JADE algorithm \cite{Cardoso1993}, and $\pi$CA\cite{Sameni2008a} using the maternal R-peaks {as reference} are shown in Figs.~\ref{fig:ICA} and \ref{fig:PiCA}, respectively. According to Fig.~\ref{fig:ICA}, JADE has extracted both the mECG and fECG; but as with all ICA methods, the order of the components is arbitrary. Fig. \ref{fig:PiCA} shows that $\pi$CA, using only the maternal R-peaks, has successfully ranked the components with respect to the mECG, but the fECG is not extracted as an independent channel (due to the low number of data channels and the closeness of the maternal-fetal subspaces of this sample data, as discussed in \cite{jamshidian2018fetal}).

The same sample data was used to evaluate the performance of the proposed NSCA algorithm. In this case, the local power envelope approach detailed in Section~\ref{sec:localenergy} was used to detect the {LPE} from channel 8 of Fig.~\ref{fig:original}. {Considering the typical fetal QRS length ($\approx$50~ms), the sliding window lengths were set to $w_1$=10~ms and $w_2$=200~ms}. The LPE detected by these window lengths can belong to either the mECG or fECG. Next, the LPE of the mECG were independently detected from abdominal channel 1 (as a channel which does not have any dominant fetal R-peaks due to {its} electrode placement). For this channel, the sliding window lengths were set to $w_1$=20~ms and $w_2$=400~ms, which are adapted for detecting the mECG segments by thresholding. The detected mECG windows were empirically expanded 15~ms from each side to include the entire {mECG} complex. Note that although the expansion of the detected mECG complex could have also been accomplished by lowering the LPE detection thresholds $\zeta_k^u$, that would increase the rate of false peaks due to background noise.

\begin{figure}[tb]
\centering
\subfigure[Reference channels and nonstationary time epochs]{
\label{fig:caseNo1NSTE}
\includegraphics[trim=.21in 0in 0in 0in,clip,width=3in, width=.47\textwidth]{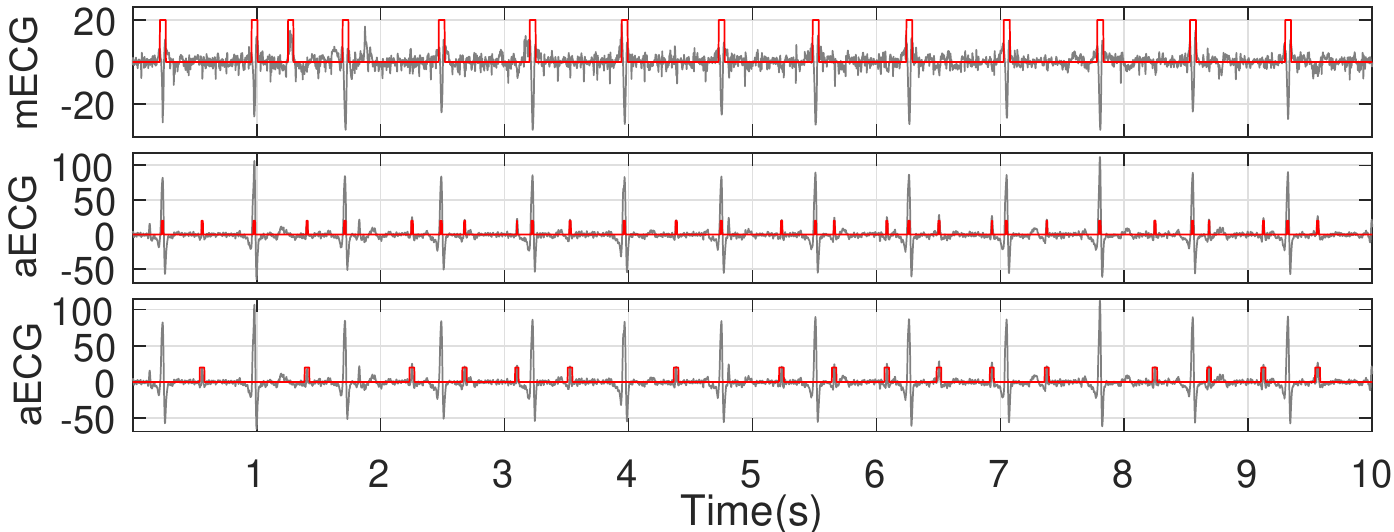}
}
\subfigure[NSCA result]{
\label{fig:caseNo1NSCA1}
\includegraphics[width=.47\textwidth]{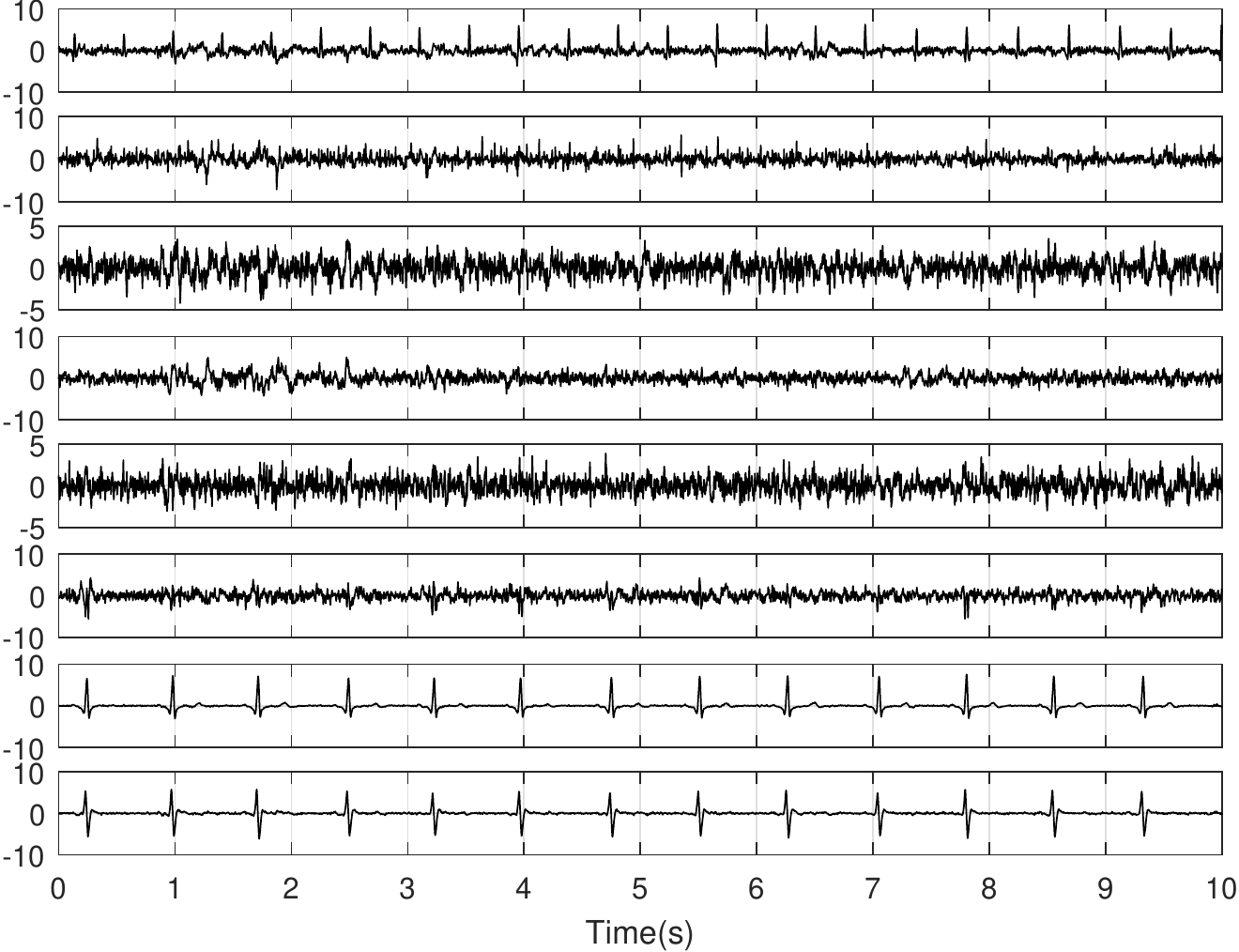}
}
\caption{The result of GEVD-based NSCA for the sample data of Fig.~\ref{fig:original}. a) the reference mECG LPE time epochs (top panel), an abdominal channel LPE time epochs (middle panel), and the merged LPE time epochs after excluding the mECG time epochs (bottom panel). The nonstationary epochs are shown as red pulses. b) The NSCA result.}
\label{fig:NSCAGEVDEnergy}
\end{figure}

Next, according to the fusion technique explained in Section~\ref{sec:fusion}, the {TNS} epochs of channel 1 were \textit{excluded} from the TNS epochs of channel 8, resulting in time instants, which mainly correspond to the fECG and not the mECG. Finally, the resulting TNS epochs were used in NSCA. The result of this method together with the detected TNS epochs are shown in Fig.~\ref{fig:NSCAGEVDEnergy}, where we can see that the fECG is successfully extracted and the components are ranked from top to bottom according to their similarity to the fECG. Furthermore, it is seen that the method is able to extract the fECG even during the temporal overlaps of the mECG and fECG, despite the fact that some of the fetal QRS peaks have not been considered among the {TNS} epochs (notice the missed fetal R-peaks at t = 1.0, 1.8, 4.0 and 4.8 seconds in the nonstationary epochs of Fig.~\ref{fig:caseNo1NSTE}). This can be explained by the fact that the required nonstationary statistics has been readily considered in $\mathbf{C}_\theta$ {from} the successfully detected fetal R-peaks.

We next visually study the performance of the proposed method in a noisy situation. In this case, we compare both the LPE and EKF-based parameters. The LPE parameters are similar to the ones used for generating the visual results, detailed above. The EKF parameters have been selected and tuned according to the procedure detailed in \cite{SSJC06,SSJ08}. For the following results, as we are only interested in the fetal QRS spikes, we empirically set the thresholds of the proposed NSCA algorithm as follows: {$\zeta_k^l= 0$, $\zeta_k^u= 3std(\rho_k(t))$, } $\mu_k = 3std(a_k(t))$, {$\lambda_k^l=0$, $\lambda_k^u = 3std(\gamma_k(t))$}, $\xi_k = 3std(q_k(t))$, $\kappa_k = 3std(\epsilon_k(t))$, where $std(\cdot)$ denotes the standard deviation. The innovation process monitoring parameters used for calculating $a_k(t)$, $\gamma_k(t)$ and $\mathbf{r}(t, \tau)$ were set to $w_a$=10~ms, $w$=10~ms and $w_r$=20~ms. Moreover, the parameter $w_a$, when used during mean removal in (\ref{eq:innovationvariance}) and (\ref{eq:innovationwhitenessindex}), was set to 50~ms. These values were empirically found by visual inspection, to be appropriate for detecting fetal QRS widths. A more rigorous approach for parameter selection would be to use prior information regarding the background and foreground signals to calculate (or to estimate numerically), the probability density functions of the indexes proposed in Section \ref{sec:nonstationaritydetection}, under the null and alternative hypothesis {and to} set the threshold to a level that fulfils the desired false alarm and detection rates. 

\begin{figure*}[tbh]
\centering
\includegraphics[trim=.05in .06in .0in .0in,clip,width=\textwidth]{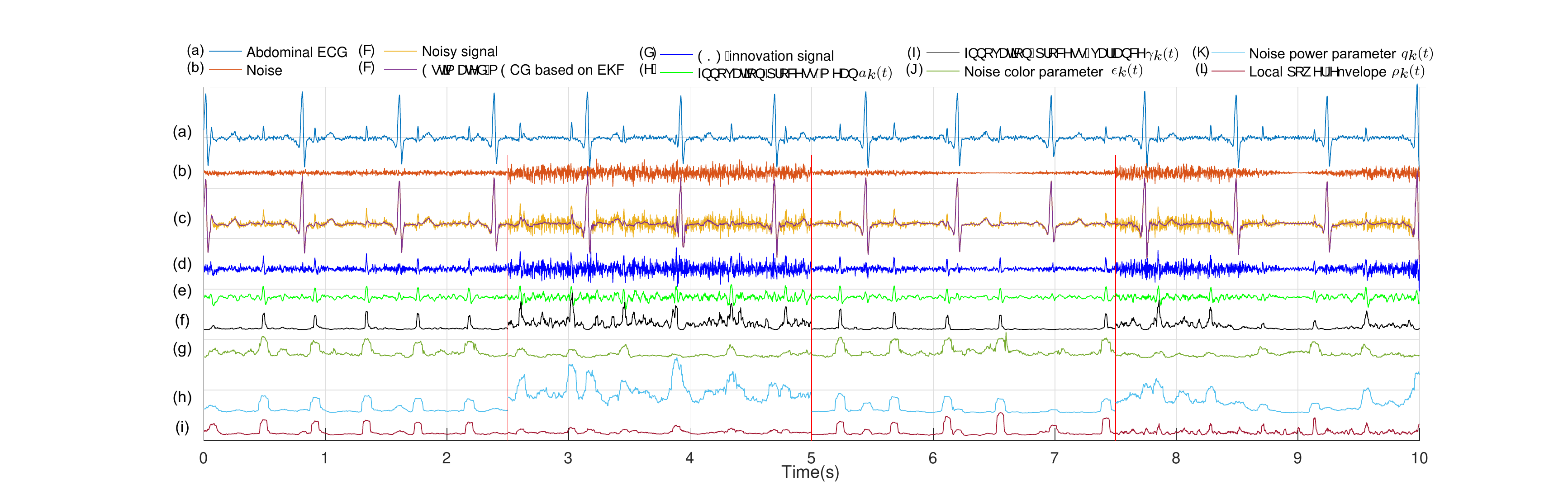}
\caption{A plot of the signal indexes used for nonstationary time epoch detection in presence of different noises. From top to bottom, the signals correspond to a) the maternal abdominal signal, b) the synthetically generated nonstationary noise, c) the estimated mECG based on EKF overlaid over the noisy signal, d) the EKF innovation signal, e) the innovation process mean $a_k(t)$, f) the innovation process variance $\gamma_k(t)$, g) the noise color parameter $\epsilon_k(t)$, h) the noise power parameter $q_k(t)$, and i) the innovation signal local power envelope $\rho_k(t)$. From left to right, the 2.5~s segments correspond to 15~dB WGN, 5~dB WGN, 15~dB NGN, and 5~dB NGN. The signals are scaled for better visualization.}
\label{fig:multiNoise}
\end{figure*}

For this example, we added white Gaussian noise (WGN) and variance nonstationary Gaussian noise (NGN) in 15~dB and 5~dB SNRs, to successive 2.5~s segments of the sample data of Fig.~\ref{fig:original}. A sample channel of the resulting noisy signal and the indexes used for calculating the TNS epochs are shown in Fig.~\ref{fig:multiNoise}. We can see how each of the indexes proposed in Section \ref{sec:nonstationaritydetection} respond to the noise type. In practice, one can choose between the proposed indexes depending on the expected background or device noise, to tailor the method for different datasets and noise scenarios, using proper thresholds.

\begin{table*}[tb]
\caption{Fetal R-peak and heart-rate detection accuracies using different AJD and GEVD schemes, in presence of different noise types and levels. The column titles refer to the indexes used for TNS detection, as defined in Sections \ref{sec:nonstationaritydetection} and \ref{sec:results}.}
 \label{tab:HRMComparison1}
 \setlength\tabcolsep{3.7pt}
\begin{center}
 \begin{tabular}{c| c |c| l l | l l l l l | l l}
 \hline
 \multicolumn{1} {p{.7cm}}{}
& \multicolumn{1} {p{.7cm}}{}
& \multicolumn{1}{p{1.2cm}|}{}
& \multicolumn{2}{p{2.4cm}|}{\centering AJD}
& \multicolumn{7}{p{8.4cm}}{\centering GEVD}
 \\
 \hline
  \multicolumn{1} {p{.7cm}|}{\centering Index}
& \multicolumn{1} {p{.7cm}|}{\centering Noise}
& \multicolumn{1} {p{1.2cm}|}{\centering SNR(dB)}
& \multicolumn{1} {p{1.2cm}}{\centering JADE}
& \multicolumn{1} {p{1.3cm}|}{\centering UWEDGE}	
& \multicolumn{1} {p{1.2cm}}{\centering $a_k(t)$ }
& \multicolumn{1} {p{1.2cm}}{\centering $\gamma_k(t)$}
& \multicolumn{1} {p{1.2cm}}{\centering $\epsilon_k(t)$  }
& \multicolumn{1} {p{1.2cm}}{\centering $q_k(t)$  }
& \multicolumn{1} {p{1.2cm}|}{\centering $\rho_k(t)$  }
& \multicolumn{1} {p{1.2cm}}{\centering GEVD-U  }
& \multicolumn{1} {p{1.2cm}}{\centering GEVD-I  }
\\ [0.5ex] 
 \hline
\multirow{10}{*}{$\text{HR}_m${(\%)}} & \multirow{5}{*}{WGN} & -5	& 43.2$\pm$34.3 &	44.2$\pm$32.4 &	46.4$\pm$32.5 &	44.5$\pm$30.1 &	\textbf{48.8}$\pm$\textbf{29.7} &	40.7$\pm$34.6 &	37.5$\pm$32.5 &	36.9$\pm$27.5 &	42.6$\pm$36.2
\\ 
& & 0 & 70.8$\pm$23.1 &	70.2$\pm$22.7 & 72.0$\pm$22.0 &	73.2$\pm$24.1 & \textbf{74.9}$\pm$\textbf{21.8} & 69.2$\pm$25.3 & 60.1$\pm$37.6 & 59.3$\pm$37.8 & 71.9$\pm$22.0
\\
& & 5 & 88.0$\pm$8.4 & 88.0$\pm$8.4 & 88.5$\pm$7.8 & 88.1$\pm$8.3 & 88.3$\pm$8.0 & 88.0$\pm$8.4 & \textbf{89.1}$\pm$\textbf{7.9} & 87.9$\pm$8.4 & 88.3$\pm$8.0
\\
& & 10 & 92.0$\pm$5.5 & 91.9$\pm$5.5 & \textbf{92.6}$\pm$\textbf{4.8} & 91.8$\pm$5.4 & 91.7$\pm$5.5 & 91.7$\pm$5.5 & 92.3$\pm$5.0 & 91.9$\pm$5.5 & 89.4$\pm$9.7
\\
& & 15 & 93.1$\pm$5.0 & 92.7$\pm$4.5 & 93.3$\pm$4.4 & 93.0$\pm$4.8 & \textbf{93.5}$\pm$\textbf{4.2} & 93.3$\pm$4.4 & 93.3$\pm$4.3 & 92.9$\pm$4.9 & 92.4$\pm$4.8
\\ \cline{2-12}
& \multirow{5}{*}{NGN} & -5 & 56.1$\pm$31.6 & 60.0$\pm$28.0 & 56.5$\pm$31.7 & 55.3$\pm$32.2 & \textbf{62.4}$\pm$\textbf{26.0} & 54.4$\pm$33.3 & 52.2$\pm$38.3 & 56.0$\pm$31.4 & 55.9$\pm$40.3%
\\
& & 0 & 80.2$\pm$16.2 & 80.1$\pm$14.1 & 81.5$\pm$14.3 & 79.1$\pm$17.0 & 82.6$\pm$14.3 & 75.7$\pm$22.4 & 81.3$\pm$14.3 & 80.2$\pm$16.2 & \textbf{83.3}$\pm$\textbf{15.7}
\\
& & 5 & 91.2$\pm$5.9 & 89.6$\pm$6.5 & 91.1$\pm$6.0 & 90.0$\pm$6.8 & 90.0$\pm$6.8 & 90.2$\pm$6.6 & \textbf{91.3}$\pm$\textbf{5.9} & 91.0$\pm$6.1 & 89.8$\pm$6.9
\\
& & 10 & 93.1$\pm$5.2 & 92.2$\pm$4.7 & \textbf{93.4$\pm$5.0}  & 92.4$\pm$5.0 & 92.6$\pm$4.9 & 92.1$\pm$5.3 & 92.9$\pm$4.8 & 93.1$\pm$5.2 & 92.2$\pm$5.6
\\
& & 15 & 93.5$\pm$4.8 & 92.7$\pm$4.3 & \textbf{94.1}$\pm$\textbf{3.8} & 93.8$\pm$4.0 & 94.0$\pm$3.9 & 93.7$\pm$4.0 & 93.6$\pm$4.1 & 93.4$\pm$4.7 & 93.6$\pm$4.1\\
 \hline 
 &  & -5 & 48.8$\pm$42.1 & 49.7$\pm$38.9 & 58.6$\pm$31.7 & 54.0$\pm$34.2 & \textbf{61.2}$\pm$\textbf{28.6} & 50.8$\pm$38.4 & 44.6$\pm$41.3 & 45.2$\pm$36.9 & 46.5$\pm$44.3
\\ 
& & 0 & 72.9$\pm$29.1 & 80.9$\pm$16.2 & 81.9$\pm$16.2 & 81.6$\pm$15.8 & \textbf{83.6}$\pm$\textbf{14.2} & 80.3$\pm$17.5 & 82.4$\pm$16.2 & 70.8$\pm$29.1 & 74.3$\pm$28.0
\\ 
& & 5 & 90.8$\pm$12.2 & 90.8$\pm$12.1 & 91.2$\pm$11.9 & 91.0$\pm$12.0 & 91.1$\pm$11.9 & 91.0$\pm$11.8 & \textbf{91.3}$\pm$\textbf{11.4} & 90.9$\pm$12.0 & 91.1$\pm$11.9
\\ 
& & 10 & 93.1$\pm$12.3 & 92.4$\pm$12.0 & \textbf{93.8}$\pm$\textbf{11.0} & 93.1$\pm$12.2 & 92.7$\pm$12.0 & 92.3$\pm$11.9 & 92.8$\pm$10.9 & 93.0$\pm$11.4 & 92.7$\pm$11.8
\\ 
& \multirow{-5}{*}{WGN} & 15 & 93.5$\pm$12.2 & 92.9$\pm$11.6 & 93.6$\pm$10.4 & 93.5$\pm$12.1 & \textbf{93.9}$\pm$\textbf{10.6} & 93.6$\pm$10.4 & 93.2$\pm$10.2 & 92.9$\pm$11.9 & 92.5$\pm$11.2
\\ \cline{2-12} 
& & -5 & 75.9$\pm$18.6 & 71.6$\pm$22.9 & \textbf{76.2}$\pm$\textbf{18.5} & 70.9$\pm$23.6 & 75.4$\pm$19.5 & 72.0$\pm$23.7 & 71.7$\pm$23.1 & 70.7$\pm$23.4 & 66.9$\pm$28.8
\\ 
& & 0 & \textbf{88.8}$\pm$\textbf{12.9} & 85.8$\pm$13.3 & 86.2$\pm$13.3 & 87.1$\pm$12.7 & 87.6$\pm$12.3 & 86.0$\pm$13.4 & 86.4$\pm$13.3 & 85.9$\pm$13.4 & 85.2$\pm$13.6
\\ 
& & 5 & 91.9$\pm$12.4 & 91.3$\pm$11.9 & 91.9$\pm$11.3 & 91.6$\pm$12.0 & 91.5$\pm$11.9 & 91.6$\pm$11.9 & \textbf{92.6}$\pm$\textbf{11.3} & 91.9$\pm$11.1 & 91.5$\pm$12.0
\\ 
& & 10 & 93.2$\pm$12.1 & 92.6$\pm$11.5 & \textbf{93.8}$\pm$\textbf{11.2} & 92.8$\pm$11.7 & 93.2$\pm$11.0 & 93.3$\pm$11.6 & 92.9$\pm$10.7 & 92.9$\pm$11.3 & 92.4$\pm$10.4
\\ 
\multirow{-10}{*}{$\text{F}_1$(\%)} & \multirow{-5}{*}{NGN} & 15 & 93.6$\pm$12.1 & 93.0$\pm$11.3 & 93.8$\pm$10.5 & 93.7$\pm$10.6 & \textbf{93.9}$\pm$\textbf{10.7} & 93.4$\pm$10.4 & 93.4$\pm$10.2 & 93.7$\pm$10.8 & 93.4$\pm$10.2
\\
\hline
\end{tabular}
\end{center}
\end {table*}

For a statistical performance analysis, the proposed and benchmark methods were applied to all records of the \textit{PhysioNet abdominal and direct fetal electrocardiogram database} \cite{MIT-BIH-ADFECG}. This dataset consists of 300~s records acquired from five different women in labor from four maternal abdominal leads and a fetal scalp lead, at a sampling rate of 1~kHz. The fetal scalp lead was only used as benchmark for performance assessment and not during source separation.

For this study, nine approaches are considered: the classical JADE algorithm used as benchmark (using the \texttt{jadeR.m} Matlab implementation by J.~F.~Cardoso \cite{CardosoSourceCodes}); GEVD-based NSCA using the five TNS epoch detection indexes proposed in Section \ref{sec:nonstationaritydetection}; GEVD using the union of the TNS epochs (GEVD-U); GEVD using the intersection of the {TNS} epochs (GEVD-I); the AJD version of NSCA using covariance matrices obtained from the EKF innovation process and five covariance matrices using the proposed TNS epoch detection indexes. AJD is performed by \textit{uniformly weighted exhaustive diagonalization with Gauss iterations (UWEDGE)} (using the \texttt{uwedge.m} Matlab implementation by P.~ Tichavsk\'y \cite{tichavsky2009fast}).

In order to quantify the performance of each method, the channels obtained after source separation were given to an automatic fECG channel selection and R-peak detection algorithm for heart-rate calculation. Note that although NSCA-based algorithms are expected to automatically rank the fECG as their first channel, there is no such guarantee for JADE or UWEDGE. Moreover, depending on the noise type and SNR, even the proposed algorithms may mistakenly rank non-fECG channels as the first channel. Another case is when more than one fECG channel is extracted or when there is a gradual fECG ``channel switching'' over time. The latter case occurs during long recordings in which the fetus moves during the data acquisition session. Therefore, we hereby do not use the automatic ranking feature of NSCA during the evaluation and give all methods the same chance by using the noted channel selection and ranking functions.

The required tools and algorithms for the post-extraction analysis and fetal heart-rate calculation were adopted from our recent work \cite{jamshidian2018fetal}, which are online available at \cite{OSET3.14}. For simplicity, a local peak detector (LPD) over a sliding window is used for post-fECG-extraction R-peak detection \cite{OSET3.14}, the signal quality indexes (SQI) proposed in \cite{jamshidian2018fetal} for automatic channel ranking, and the percentage of $\text{HR}_m$ and $\text{F}_1$ measures detailed in \cite{behar2014comparison} are used to evaluate the different methods. The $\text{HR}_m$ measure is the ratio of the number of reference fetal HRs to the number of detected fetal HRs that are within $\pm$5 beats per minute of their corresponding reference measurement. The fetal HR is obtained from the output of each separation method, while the reference fetal HR is obtained from the scalp lead. It is a continuous index ranging between 0 and 1 (reported in percentage in the sequel), corresponding to the worst and best fetal heart-rate detection performances, respectively. The $\text{F}_1$ index measures the performance of fetal R-peak detection within an acceptable tolerance window (typically 50~ms for fECG) around the reference fetal R-peak. Further details regarding the utilized performance measures can be found in \cite{behar2016practical}.

Since the objective is to assess the fECG extraction performance, no post-processing was performed on the extracted fetal heart-rates to correct the missing or wrongly detected fetal R-peaks, before {calculating the performance indexes}.

Table \ref{tab:HRMComparison1} summarizes the average and standard deviation of $\text{HR}_m$ and $\text{F}_1$ over the best fECG channel after fECG extraction (selected automatically), for WGN and NGN in the SNR range of -5~dB to 15~dB (per channel), in 5~dB steps, using distinct synthetic noise added to each channel of the records. Accordingly, the performance of all methods is better under nonstationary noises, as compared to white noise. In all methods, the standard deviations of the results are clearly very high in zero and negative SNRs, which is associated to the fact that fetal R-peak detection becomes more erroneous in these SNR. Nevertheless, as compared with JADE and AJD, the hereby proposed methods based on TNS have been more robust (in terms of mean and standard deviation). In low SNR (5~dB and below) GEVD-based methods that use the innovation process spectral color parameter $\epsilon_k(t)$ and the innovation process mean parameter $a_k(t)$ show the best performances. After these two features, the union of all TNS epochs has the most robust performance among the GEVD-based methods. JADE and AJD using UWEDGE show average performance in all SNR. However, UWEDGE using covariance matrices obtained from five nonstationarity measures always outperforms JADE, which proves the hypothesis that approximate joint diagonalization, or even the exact diagonalization of well-targeted matrices is preferred over totally blind methods (whenever possible).

\section{Discussion and conclusion}
\label{sec:conclusion}
In this paper, an algorithm was proposed for the separation (or extraction) of temporally nonstationary signals, using a set of indexes for detecting the nonstationarity time epochs. It was shown that by monitoring the local power variations and the mean, variance and spectral color of an extended Kalman filter's innovation process, it is possible to identity epochs of nonstationarity, which are later used to calculate targeted covariance matrices used for joint diagonalization. It was also shown how the nonstationarity time labels can be {merged} together to obtain more robust labels using multiple indexes and channels. The GEVD-based version of the algorithm requires accurate covariance matrices, but has the advantage of a higher performance and the ability of sorting the components in order of similarity to the TNS epochs. On the other hand, the AJD version is more robust to the selection of the matrices, but does not guarantee the order of extracted sources.

Various aspects of this study can be extended in future research, including:
\begin{itemize}[leftmargin=*]
    \item Temporally nonstationary components can be rather sparse in time. Therefore, the statistics used to estimate the sample covariance matrices can be insufficient; resulting in inaccurate covariance and separation matrices. Using recent developments in \textit{random matrix theory} \cite{engle2017large}, the performance of GEVD-based methods can be significantly improved by more accurate estimation of the generalized eigen-matrix used for source separation.
    \item The detection and fusion of the nonstationarity labels can be made more systematically, using data-driven priors.
    \item For low-rank and degenerate mixtures, the proposed method can be integrated into the deflation algorithm proposed in \cite{SJS2010}, specifically for fECG extraction.
    \item By using incremental updates of the covariance matrices in combination with source separation schemes such as \textit{equivariant adaptive separation via independence} (EASI) \cite{Cardoso1996}, online versions of the proposed method can be obtained.
    \item Apart from the studied application, the proposed method has many other applications including single-trial event-related potential detection and extraction{, or the extraction of spatially nonstationary events in medical images.}
    \item As detailed in Section \ref{sec:NSCA}, the idea of nonstationarity event identification/extraction can be more generally perceived as a means of detecting occasional model mismatches, spatio-temporal anomalies or rare samples, which is a recurrent problem in machine learning applications such as clustering and classification. This aspect of the proposed framework merits further studies in future research.
\end{itemize}

\section*{Acknowledgement}
This work has been partially funded by the European Research Council Advanced Grant Number 320684, on Challenges in the Extraction and Separation of Sources (CHESS).
\bibliographystyle{IEEEtran}
\bibliography{IEEEabrv,References}
\end{document}